\documentstyle[prl,aps,epsf]{revtex}

\newcommand{\A}{{\bf A}}
\newcommand{\D}{{\bf D}}

\newcommand{\n}{{\bf n}}
\newcommand{\Eq}[1]{Eq.~(\ref{#1})}
\newcommand{\Fig}[1]{Fig.~\ref{#1}}
\newcommand{\Figure}[1]{Figure~\ref{#1}}

\begin{document}
\advance\textheight by 0.2in
\draft
\twocolumn[\hsize\textwidth\columnwidth\hsize\csname@twocolumnfalse\endcsname

\title{Critical exponent $\eta_\phi$ of the Lattice London
  Superconductor\\ and vortex loops in the 3D XY model}

\author{Peter Olsson}

\address{Department of Theoretical Physics, Ume\aa\ University, 
  901 87 Ume{\aa}, Sweden}

\date{\today}   

\maketitle

\begin{abstract}
  The anomalous dimension of the lattice London superconductor is
  determined from finite size scaling of the susceptibility to be
  $\eta_\phi = -0.79(1)$. Indirect determinations of $\eta_\phi$
  from properties of the vortex loops in the 3D XY model are also
  attempted, but it is found that the results are sensitive to details
  in the simulations related to vortex loop intersections.  It is
  suggested that the same value of $\eta_\phi$ can at most be obtained
  from vortex loop properties in the limit of low vortex density.
\end{abstract}


]

The properties of the Meissner transition is a classical problem in
statistical physics. Whereas this transition was originally believed
to always be of first order, the work by Dasgupta and
Halperin\cite{Dasgupta_Halperin} gave strong arguments that the
transition should instead be continuous.  The argument is based on a
duality transformation of the lattice London Superconductor (LLS) and
suggests that the transition should be 3D XY-like, but with the
temperature scale inverted.  A direct consequence of this relation is
the expectation that the correlation length exponent $\nu$ should be
the same in the superconductor as in the 3D XY (planar rotor) model.
However, with fluctuations in both the phase angle and the gauge
field, it becomes possible to define two characteristic lengths, and
the behavior of the magnetic screening length $\lambda$ has recently
been a subject of some controversy\cite{Kiometzis_KS}. The current
evidence\cite{Olsson_Teitel:vi3l,Hove_Sudbo} points to a scenario
where both characteristic lengths diverge with the same exponent. This
is related to the presence of an anomalous dimension $\eta_A = 1$ for
the gauge fluctuations\cite{Herbut_Tesanovic}.

What is needed beside $\nu$ and $\eta_A$ for the characterization of
the universality class of the transition is a knowledge of the
anomalous dimension $\eta_\phi$ associated with the phase
correlations.  This quantity was first determined to be $\eta_\phi =
-0.20$ from a renormalization group calculation to one loop
order\cite{Herbut_Tesanovic}.  Determinations from simulations have so far
only been indirect through the properties of vortex loops in the 3D XY
model\cite{Nguyen_Sudbo:99}. The main motivation for that analysis was
the possible connection between the sign of $\eta_\phi$ and the
existence of a vortex loop blowout transition in high-$T_c$
superconductors in applied magnetic
fields\cite{Tesanovic:99,Nguyen_Sudbo:99}.

In this Letter we report on a direct determination of the anomalous
dimension $\eta_\phi$ from Monte Carlo (MC) simulations of the LLS that
gives a surprisingly large negative value, $\eta_\phi = -0.79\pm
0.01$.  We also consider in detail the approach to determine this
exponent from the properties of the vortex loops in the 3D XY model.
The main conclusion from that study is that the results depend
strongly on the details in the simulation related to vortex loop
intersections. We suggest that the theoretically expected value can at
most be found in the limit of low vortex density.

The Hamiltonian of the LLS\cite{Dasgupta_Halperin,Olsson_Teitel:vi3l} is
\begin{equation}
  H = \sum_{i\mu} \left\{U(\phi_{i+\hat{\mu}} - \phi_i - A_{i\mu}) +
  \frac{1}{2} J \lambda_0^2 [\D \times \A]^2_{i\mu} \right\},
  \label{Hamilt}
\end{equation}
where $\phi_i$ is the phase of the superconducting wave function on
site $i$, and $A_{i\mu}$ is the vector potential on the link starting
at site $i$ and pointing along $\hat{\mu}$.  The sum is over all bonds
of a 3D simple cubic lattice of unit grid spacing and $\mu = x$, $y$,
$z$. In the first term, the kinetic energy of flowing supercurrents,
$U(\varphi)$ is the Villain function\cite{Villain:75}
\begin{displaymath}
  e^{-U(\varphi)/T} = \sum_{p=-\infty}^\infty e^{-J(\varphi-2\pi p)^2/(2T)}.
\end{displaymath}
In the second term, the magnetic field energy, $\lambda_0$ is the
bare magnetic penetration length, and $\D\times\A$ is the discrete
circulation of the vector potential.
% The coupling $J$ is related to $\lambda_0$ and the flux quantum
% $\phi_0$ through $J = \phi_0^2 / 16\pi^3 \lambda_0^2$.

In our simulations we fix $\lambda_0 = 0.3$ to be able to compare with
Ref.\ \onlinecite{Olsson_Teitel:vi3l} and fix the gauge through
$\D\cdot\A = 0$ to facilitate a simple determination of the spin
correlations.  To fulfill this constraint we update the $A_{i\mu}$ by
simultaneously adding $\delta A$ to four $A_{i\mu}$ around an
elementary plaquette, a procedure which was also used
in Ref.\ \cite{Hove_Mo_Sudbo}.  We perform our MC simulations with
the standard Metropolis algorithm, first sweeping sequentially through
the phase angles $\theta_i$ and then sweeping three times with
attempts to change the circulation of $\A$ in the three different
directions. The length of the runs close to $T_c$ were at least $10^7$
sweeps through the lattice, but often much longer.
The measured quantities discussed here are the second and fourth
moment of the magnetization,
\begin{displaymath}
  m^p = \left|\frac{1}{L^3}\sum_i e^{i\phi_i}\right|^p,
\end{displaymath}
which are used to obtain the dimensionless fraction
\begin{equation}
  Q = \frac{\langle m^2 \rangle^2}{\langle m^4\rangle},
  \label{Eq-Q}
\end{equation}
similar to Binder's cumulant, which is used to determine $T_c$.  The
susceptibility at the transition is obtained from the standard
relation $\chi = L^3\langle m^2\rangle$.

\begin{figure}
  \epsfxsize=8.2truecm
  \epsffile{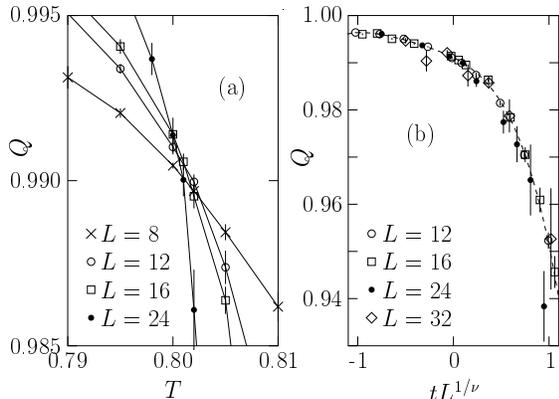}
  \caption{Determination of $T_c$ from the crossing of $Q$ given by
  Eq.~(\protect\ref{Eq-Q}). Panel (a) shows data for $L=8$ through 24;
  the $L=8$ data lies below the crossing point. Panel (b) shows the
  scaling collapse for sizes $L = 12$ through 32.}
  \label{fig-bind}
\end{figure}

The critical temperature for the LLS with the same parameter,
$\lambda_0 = 0.3$, has already been determined with high
precision\cite{Olsson_Teitel:vi3l}. Our present analysis of $Q$
therefore mainly serves to confirm that the critical properties may
be correctly determined from our simulations with $\D\cdot\A = 0$.
\Figure{fig-bind}a shows $Q$ versus temperature for system sizes $L=8$,
12, 16, and 24. We find that the curves for $L>8$ cross at $T\approx
0.80$.  The $L=8$ data does however not cross the other curves at the
same temperature, a typical sign of corrections to scaling. This data
is therefore not included in the scaling collapse.
To get a more precise determination of $T_c$ we use all our data for
$12\leq L \leq 32$ in a narrow range around $T_c$, $|tL^{1/\nu}| <
1.1$, and assume the scaling form $Q(t,L) = f_Q(tL^{1/\nu})$, where $t
= T/T_c - 1$, and $f_Q(x)$ is a scaling function.  We fix $\nu =
0.672$\cite{Hasenbusch_Torok} and let $f_Q(x)$ be a fifth order
polynomial in $x$. The data collapse nicely, \Fig{fig-bind}b, and we
find $T_c = 0.800\pm 0.001$ in good agreement with $T_c = 0.8000\pm
0.0002$ from Ref.\ \cite{Olsson_Teitel:vi3l}.

\begin{figure}
  \epsfxsize=8truecm
  \epsffile{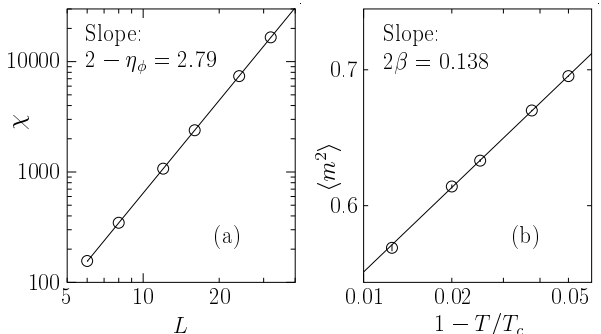}
  \caption{Panel (a) shows a finite size scaling of the susceptibility at $T_c =
    0.8$. The data fits nicely to a line with slope $2-\eta_\phi =
    2.79$. Panel (b) shows the determination of the exponent $\beta$
    from the $T$-dependence of $\langle m^2\rangle$ for $L=64$.}
  \label{fig-eta-beta}
\end{figure}

To determine the anomalous dimension $\eta_\phi$ we make use of the
standard finite size scaling relation $\chi(T_c,L) \sim L^{2-\eta_\phi}$.
The susceptibility at $T_c=0.8$ is shown in \Fig{fig-eta-beta}a. The
points do indeed to an excellent approximation fall on a straight line
and our result is $\eta_\phi = -0.79\pm 0.01$.  Panel (b) shows a
determination of the exponent $\beta$ from the temperature dependence
of $\langle m^2\rangle$. This data is for temperatures at which we
expect the finite size effects to be negligable. The result is $\beta
= 0.069\pm0.003$ in good agreement with what we expect from
$\eta_\phi$: $\beta = \frac{1}{2}\nu(d-2+\eta_\phi) = 0.070\pm 0.003$.

We now turn to the alternative and indirect approach to determine
$\eta_\phi$ through the behavior of the dual model which is the
ordinary 3D XY model\cite{Dasgupta_Halperin}. The
idea\cite{Nguyen_Sudbo:99} is that the vanishing of the line tension
of the vortex loops (see below) as $T\rightarrow T_c^-$ is governed by
$\gamma_\phi$,
\begin{equation}
  \epsilon \sim (1 - T/T_c)^{\gamma_\phi},
  \label{gamma-epsilon}
\end{equation}
which is related to $\eta_\phi$ through the Fisher scaling relation
$\gamma_\phi =\nu(2-\eta_\phi)$. The basis for \Eq{gamma-epsilon} is
the connection between $\phi$ correlations in the LLS and the vortex
loops in the 3D XY model\cite{Nguyen_Sudbo:99}. Whereas this less
direct determination of $\eta_\phi$ of course is more prone to errors,
it is interesting to examine if the same value of $\eta_\phi$ can be
obtained in that way. We will in the following examine the vanishing
of $\epsilon$ for two different cases, examine the distribution of
vortex loop diameters at $T_c$, and finally comment on the analysis in
Ref.\ \cite{Nguyen_Sudbo:99} that gave $\eta_\phi \approx -0.18$.

The standard way to locate vortices in a XY model is from the angular
difference $\varphi_{ij} = \theta_i - \theta_j$ between nearest
neighbors, with $|\varphi_{ij}| < \pi$. For each plaquette the
vorticity is then obtained from
\begin{equation}
  n = \frac{1}{2\pi} (\varphi_{12} + \varphi_{23} + \varphi_{34} + \varphi_{41}).
  \label{n-vort}
\end{equation}
With a Villain interaction it is also possible to define vortices in a
different way, which is inspired by the wire model\cite{Vallat_Beck}.
For each link one first calculates $\varphi_{ij}^0 = \theta_i -
\theta_j$ and then probabilistically sets the angular difference to
$\varphi_{ij} = \varphi_{ij}^0 - 2 \pi p_{ij}$, where $p_{ij}$ is an
integer chosen with the relative weight $e^{-J(\varphi_{ij}^0-2\pi
  p_{ij})^2/(2T)}$.  The vorticity is finally calculated from
\Eq{n-vort}.  The vortices obtained in that way are exactly equivalent
to the dual vortex line model, a fact that follows from a
straightforward generalization of the results in Ref.\ 
\cite{Vallat_Beck} to three dimensions.  The density of these
probabilistic vortices (PV) is always higher then the density of
standard vortices (SV).

The simulations of the 3D XY model were performed with the Wolff
cluster update method and Villain interaction on a cubic lattice with
$L=128$. The average for each data point is based on typically $10^5$
measurements.  The simulations were performed with both SV and PV
discussed above.  The tracing out of vortex loops were done by always
choosing the path randomly when two (or more) vortex loops meet at the
same elementary cube.

After identifying the vortex loops we measure both the perimeter $p$
and the ``diameter'' $a$ of each such loop. Whereas $p$ is directly
given by the number of vortex loop segments the calculation of $a$ is
somewhat more involved. With a loop made up by $p$ segments with
each segment starting at ${\bf r}_i$ and pointing along the unit vector
$\n_i$. the equivalent to the ``magnetic moment'' is
\begin{displaymath}
  {\cal M} = \sum_{i=1}^p {\bf r}_i \times \n_i,
\end{displaymath}
and the ``diameter'' becomes $a = 2 \sqrt{|{\cal M}|/\pi}$.  From these
values we calculate the distributions $D(p)$ and $D(a)$.

\begin{figure}
  \epsfxsize=8.2truecm
  \epsffile{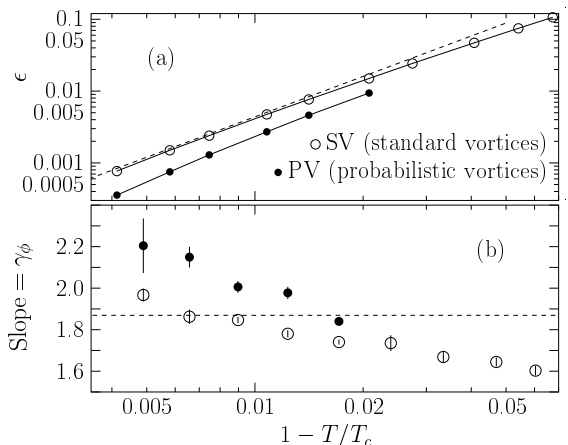}
  \caption{Panel (a) shows the line tension versus reduced temperature
    from our analyses of the vortex loops. The open circles (SV data)
    seem to agree reasonable well with the lower dashed line with
    slope $\gamma_\phi = 1.87$. The solid circles are obtained with a
    different vortex definition (PV) that gives a higher vortex
    density.  Panel (b) gives the local slope determined from
    neigboring points in panel (a).}
  \label{fig-epsilon}
\end{figure}

The line tension $\epsilon$ is obtained by
fitting\cite{Nguyen_Sudbo:99}
\begin{equation}
  D(p) \propto p^{-\alpha} \exp(-\epsilon p/T),
  \label{eq-Dp}
\end{equation}
with $\alpha$, $\epsilon$ and a prefactor as free parameters. To get
good quality fits one has to restrict the fitting to large $p$ only.
We found that good quality fits are obtained by only using data for $p
> T/\epsilon$.  \Figure{fig-epsilon} shows the line tension $\epsilon$
determined with both SV and PV (open and solid symbols, respectively).
Note that we are here examining a very narrow temperature region to be
able the probe the behavior in the critical region; most data points
are within a few percent below $T_c=3.0024(1)$\cite{Olsson:3DXY-crit}.
With logarithmic scales on both axes the slope as $T\rightarrow T_c$
is expected to give the exponent $\gamma_\phi$. To facilitate a
comparison with the above obtained $\eta_\phi = -0.79$ we draw a
dashed line in \Fig{fig-epsilon}a that corresponds to $\gamma_\phi =
\nu(2-\eta_\phi) = 1.87$. The slopes of the data seem to be in
reasonable agreement with that prediction.

A more detailed analysis, however, casts doubt on this agreement.
Panel (b) in \Fig{fig-epsilon} shows the temperature dependence of the
local slope where each value is calculated from two consecutive
points. The curvature in the original data is here seen as a trend
towards larger values as $T_c$ is approached. There is, however, no
evidence that the local slope saturates at $\gamma_\phi = 1.87$
(horizontal dashed line); the last point lies significantly above the
predicted value. To check for finite size effects the relevant points
in panel (a) have also been analyzed with $L=192$ but with no change
in the result.  For the PV data (solid circles) this deviation is even
more significant.

We will below present more evidence that suggest that, generally
speaking, the properties of vortex loops at or close to $T_c$ tend to
not agree with theoretical predictions. We will also argue that this
discrepancy is related to the existence of vortex line intersections
which are normally not accounted for in theoretical considerations.
The quantity we focus on is $D(a)$ -- the distribution of vortex loop
diameters -- at $T_c$. There is a simple analytical prediction for
this quantity based on the scale invariance of the system at
$T_c$\cite{Vachaspati_Vilenkin},
\begin{displaymath}
  D(a) \propto a^{-\lambda},\quad\lambda=4\mbox{\ at\ } T_c.
\end{displaymath}
Assume that there is on the average $N$ loops with diameter in the
interval $(a,a + da)$ in a system with volume $L^3$, i.e.\ $D(a)da =
N/L^3$. Scale invariance implies that the distribution should not be
changed by a rescaling of the system.  After rescaling with a factor
$s$ the new system size is given by $L' = L/s$ and $N$ is now the
number of loops with diameter in the interval $(a/s, a/s+da/s)$.  We
therefore get $D(a/s)da = s N/(L')^3 = s^4 D(a)da$ which in turn leads
to $D(a) \propto a^{-4}$.

\begin{figure}
  \epsfxsize=8.0truecm
  \epsffile{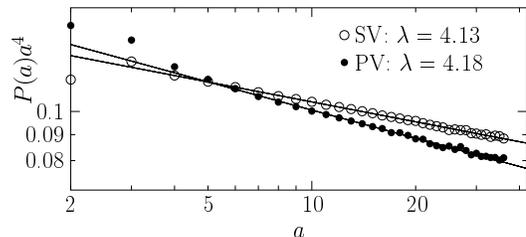}
  \caption{Distribution of loop diameters at
    $T=3.002 \approx T_c$. The expected $\lambda=4$ would give a
    horizontal line.}
  \label{fig-ahist-tw}
\end{figure}
As shown in \Fig{fig-ahist-tw} this expectation for $\lambda$ is not
confirmed by the simulations\cite{FTBC}. The values of $\lambda$ for
SV and PV are instead found to be $4.13(1)$ and $4.18(1)$,
respectively. The deviation from the expected $\lambda=4$ increases
with increasing vortex density and appears to be proportional
to the vortex density, $0.156$ and $0.226$, respectively.

\begin{figure}
  \epsfxsize=8.0truecm
  \epsffile{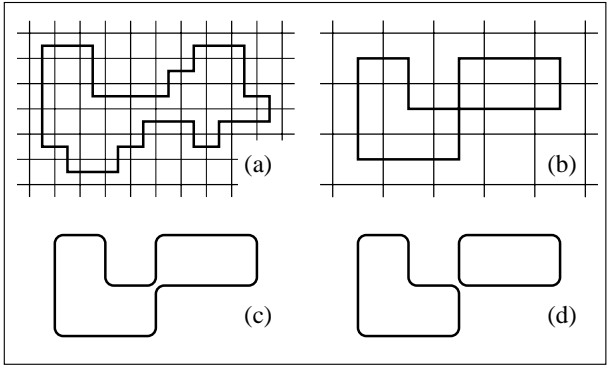}
  \caption{The rescaling of a vortex loop (for simplicity shown as a
  coarsening of the vortex lattice) may give rise to new vortex line
  intersections. This is suggested as the reason why the analytically
  expected $\lambda=4$ is not found in the simulations.}
  \label{fig-loop}
\end{figure}

We now suggest that the reason for $\lambda\neq 4$ is related to the
existence of vortex loop intersections.  Consider the rescaling of a
loop with a scale factor $s$. A non-trivial effect of the rescaling is
related to the smaller resolution of a rescaled loop. This may lead to
new vortex intersections that could change the outcome of the tracing.
As an example, consider the vortex line loop in \Fig{fig-loop}. The
original loop (a) will always be treated as a single one, but since
the rescaled loop (b) enters the same elementary cube twice the
tracing can now give the loops in panel (c) or (d), each with a
probability of 50\%. Since a splitting will give an increased weight
to smaller loops, the effect is to make $\lambda$ larger. Note that a
higher vortex density will make this effect more important which is in
accordance with our results for $\lambda$.

From the above it seems that the presence of vortex loop intersections
is the reason for the failure to obtain the expected $\lambda=4$ from
$D(a)$. One could expect the vortex loop intersections to also affect
other vortex loops properties. This is in accordance with our results
from the analysis of the vortex loop line tension that the high
density data (PV) is far off whereas the low density data (SV) is
fairly close to the expected behavior. The correct behavior of
$\epsilon$ should then at most be expected in the limit of low vortex
density.

\begin{figure}
  \epsfxsize=8truecm
  \epsffile{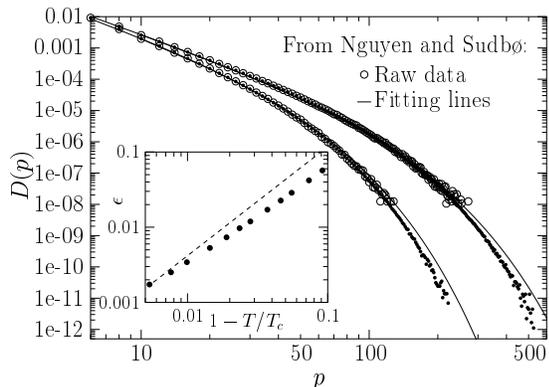}
  \caption{$D(p)$ for the 3D XY model with cosine interaction and
    splitting of self-intersecting loops.  The fitting lines from
    \protect\cite{Nguyen_Sudbo:99} do not agree well with our more
    precise data (solid dots). The inset shows our values for
    $\epsilon$; the slope is clearly different from $\gamma_\phi=1.45$
    (dashed line) obtained in Ref.\ \protect\cite{Nguyen_Sudbo:99}.}
  \label{fig-phist-saw}
\end{figure}

The determination of the line tension for vortex loops in a 3D XY
model was first made in Ref.\ \cite{Nguyen_Sudbo:99}. Their
simulations were with cosine spin interaction ($T_c\approx 2.2$) and a
different method for tracing out the loops; self-intersecting vortex
loops are always split into two. They found $\eta_\phi = -0.18\pm
0.07$, in agreement with the analytically obtained $\eta_\phi\approx
-0.2$\cite{Herbut_Tesanovic}.  However, a comparison with our more
precise MC data (now obtained with cosine interaction and splitting of
intersecting loops) casts strong doubt on their analysis.
\Figure{fig-phist-saw} shows our values for $D(p)$ (dots) together
with both the data (open circles) and the fitting curves (solid lines)
for $T = 2.0$ and 2.1 from Ref.\ \cite{Nguyen_Sudbo:99}. We first note
that the two different sets of MC data agree well.  From the large-$p$
part of the data it is however clear that the fitting curves from
Ref.\ \cite{Nguyen_Sudbo:99} do not agree with our data. The same
discrepancy can actually also be seen for $T = 2.0$ in Fig.\ 3 of
Ref.\ \cite{Nguyen_Sudbo:99}.
% The same (incorrect) exponent
% ($\gamma_\phi \approx 1.45 \Rightarrow \eta_\phi = -0.18$) may be
% obtained from our data by including all data with $D(p) < 10^{-4}$ in
% the fit to \Eq{eq-Dp}; though at the price of a very bad quality of
% the fit.
% The inset in \Fig{fig-phist-saw} shows our values for
% $\epsilon$ obtained by only including data for $p>1.2 T/\epsilon$.
The inset in \Fig{fig-phist-saw} shows our values for $\epsilon$
obtained from a fit to \Eq{eq-Dp} only including data for $p>1.2
T/\epsilon$.  The slope differs clearly from $\gamma_\phi = 1.45$
shown by the dashed line.  We therefore conclude that the good
agreement\cite{Nguyen_Sudbo:99} with the analytically obtained
$\eta_\phi\approx -0.2$ was only accidental.  In our analysis the
slope approaches $\gamma_\phi\approx1.15$ as $T\rightarrow T_c$. If
one instead allows for vortex loop intersections $\gamma_\phi$ behaves
much the same as in \Fig{fig-epsilon} (again with $\gamma_\phi$
signficantly above 1.87).  This again shows the central role of the
intersections for the vortex loop properties.

To conclude, our main result is a direct determination of the anomalous
dimension in the LLS, giving $\eta_\phi = -0.79\pm 0.01$. We have also
attempted determinations of the same quantity from the vanishing of
the line tension in the 3D XY model, but found that it was not
possible to get a well-defined value for the exponent $\gamma_\phi$.
From considerations involving the distribution of vortex loop
diameters we suggest that it is the vortex loop intersections that are
responsible for the failure to find the analytically expected values
of some vortex-loop related exponents. This implies that the
``correct'' behavior at criticality can at most be expected in the
limit of low vortex density.

The author thanks P.~Minnhagen, A.~Sudb\o, S.~Teitel, and
Z.~Te\u{s}anovi\'{c} for valuable discussions. This work was supported
by the Swedish Natural Science Research Council through Contract No.\ 
E 5106-1643/1999.

\bibliographystyle{prsty} 
%\footnotesize 
%\bibliography{lett,../o}

\end{document}